\pdfoutput=1
\documentclass[journal]{IEEEtran}
\usepackage[T1]{fontenc}
\usepackage[utf8]{inputenc}
\usepackage{amsmath,amssymb}
\usepackage{graphicx}
\usepackage{booktabs}
\usepackage{array}
\usepackage{makecell}
\usepackage{cite}
\usepackage{url}
\usepackage[hidelinks]{hyperref}
\newcommand{\cit}[3]{\begin{tabular}[c]{@{}c@{}}$#1$\\[-2pt]{\scriptsize$[#2,\,#3]$}\end{tabular}}
\begin{document}
\title{What Standard Evaluations Hide About Detectors of Hijacked LLM Agents: Rare Attacks, Unseen Tools and Benchmark Transfer}
\author{Mahmoud Abbasi,~\IEEEmembership{Member,~IEEE}, Juan M. Núñez V., and Amin Shahraki,~\IEEEmembership{Senior Member,~IEEE}%
\thanks{M. Abbasi (corresponding author, e-mail: mahmoudabbasi@usal.es) is with the University of Salamanca, 37007 Salamanca, Spain.}%
\thanks{J. M. Núñez V. is with the AIR Institute, Salamanca, Spain.}%
\thanks{A. Shahraki is with the Department of Informatics, University of Oslo, 0373 Oslo, Norway.}}
\markboth{Preprint}{Abbasi \MakeLowercase{\textit{et al.}}: What Standard Evaluations Hide About Detectors of Hijacked LLM Agents}
\maketitle

\begin{abstract}
Detectors can flag hijacked large language model (LLM) agents from the traces of the tools they call. In the eleven evaluations we audited, attacks or anomalies make up 16\% to 86\% of the test samples, no evaluation varies that share, and six report no test on data from a second benchmark or source. We ask what such evaluations hide. On the AgentDrift and AgentDojo benchmarks we test trained detectors that read tool identities, behavior, call sequences, step embeddings or text, and a released LLM action monitor, AlignmentCheck, with thresholds fixed on training data or set by the monitor's released decision rule. First, at a 2\% attack rate a detector with an area under the receiver operating characteristic curve (AUROC) of 0.879 reaches a precision of 0.076, and the mix of negatives alone moves that precision between 0.039 and 0.117. Second, as tool coverage, the share of test tool calls whose tool was seen in training, falls, detector rankings cross near 50\% coverage and reverse below it, because unseen tools map to inputs that training never reached. Dropout on tool inputs largely removes the reversal. Third, no detector trained on AgentDrift ranks AgentDojo traces better than chance (AUROC 0.47 to 0.52), including one that reaches 0.996 within AgentDojo. The monitor, trained on neither, reaches 0.930, but with its open backbone its released decision rule would still mostly raise false alerts at a 2\% attack rate (precision 0.096). We turn these findings into a checklist for evaluating trace detectors.
\end{abstract}


\section{Introduction}

Agents built on large language models (LLMs) carry out user tasks by calling tools such as mail clients, payment services, file systems and web browsers, and the output of each call returns to the model as context. Instructions planted in that output can redirect the agent toward an attacker's goal, a threat known as indirect prompt injection \cite{greshake2023}. Benchmarks such as InjecAgent \cite{zhan2024injecagent} and AgentDojo \cite{agentdojo} show that such injections still succeed against current models at rates that cannot be ignored. A hijacked agent has to act through its tools, so the record of its tool calls is a natural place to look for evidence of compromise. Recent work proposes rule, automaton, graph, sequence and LLM-based detectors that score such traces \cite{traceaegis,praetor,agentsentry,driftnet,trajguard,mcpshield,skynet,trajad,sentinelagent,llamafirewall}.

How these detectors are evaluated matters as much as how they are built. In the eleven evaluations we audited (Table~\ref{tab:audit}), test sets that mix both classes contain 16\% to 86\% attacks or anomalies, only four works report a threshold-free measure, none varies the attack share, and six report no cross-benchmark or within-target test. A security team that audits agent sessions meets different conditions. Successful hijacks are likely to be a small and unknown share of its sessions, and the tools, tasks and domains its agents meet change over time and between organizations.

Three consequences are known in general but have not been measured for these detectors. When attacks are rare, even a small false-positive rate yields alerts that are mostly false \cite{axelsson2000}, and the receiver operating characteristic (ROC) hides this because, unlike precision, it does not depend on prevalence \cite{davis2006,saito2015}. Learned security detectors degrade under distribution shift and profit from biased splits \cite{sommer2010,pendlebury2019,arp2022}. And a model can learn a cue that separates the classes in one dataset without defining the behavior of interest \cite{geirhos2020}. Tool-call detectors face one further shift that is easy to measure, the share of test tool calls whose tool appeared in training, which we call tool coverage.

Parts of this picture have been examined. MCPShield \cite{mcpshield} shows that random splits can inflate AUROC by up to 25.8 points relative to task-disjoint splits and traces the gap to metadata features, a one-hot tool name with an argument hash and the response length, that learn task-specific tool signatures. Agent-Sentry \cite{agentsentry} adds a within-target control, TrajAD \cite{trajad} holds out a domain, and Fomin \cite{fomin2026} makes the leave-one-dataset-out case for classifiers of malicious prompts. No audited evaluation, however, measures detection at controlled attack prevalence, reports tool coverage for every split, holds out every test tool, or transfers from a synthetic benchmark to recorded agent runs.

\begin{figure*}[t]
\centering
\includegraphics[width=\textwidth]{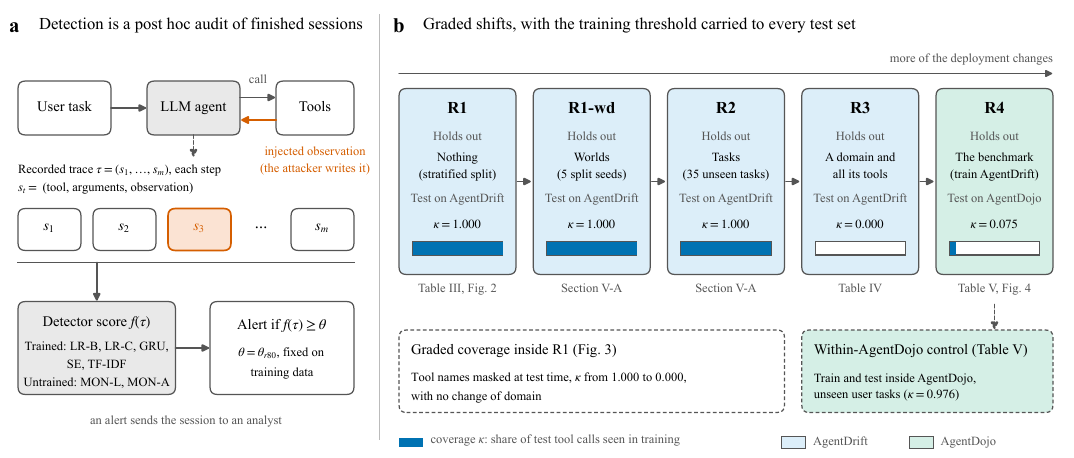}
\caption{Detection setting and evaluation design. (a) The attacker controls the content of a tool observation, and the detector scores the recorded trace of each finished session against a threshold fixed on training data. (b) The settings of Section~\ref{sec:settings}, from the standard split R1 to cross-benchmark transfer R4, with the tool coverage $\kappa$ of each test set (the share of its tool calls whose tool occurs in training) and where each is reported. Dashed boxes are the two controls, graded coverage inside R1 and the within-AgentDojo control. Thresholds are never re-tuned on test data.}
\label{fig:overview}
\end{figure*}

We run these tests jointly on one set of detectors, with thresholds always fixed on training data, and ask three questions (Fig.~\ref{fig:overview}).
\begin{itemize}
\item \emph{Q1, rare attacks.} What does a high AUROC say about the alerts a detector raises when 1\% to 10\% of sessions are attacked?
\item \emph{Q2, unseen tools.} What happens to detector rankings and thresholds as the share of known tools falls, and why?
\item \emph{Q3, a new benchmark.} Does a detector trained on one benchmark keep its signal on another?
\end{itemize}
We study simple trained detectors on tool identity, behavior and call sequences, a reimplementation of a published step-embedding architecture, a content reference, and a released LLM action monitor, AlignmentCheck \cite{llamafirewall}, with an open and a commercial backbone. The answers are as follows.
\begin{itemize}
\item At a 2\% attack rate, the combined logistic regression has an AUROC of 0.879 but a precision of 0.076 at its training threshold, 12.2 false alerts for every true one, and the share of hard negatives alone moves this precision between 0.039 and 0.117 (Section~\ref{sec:rare}).
\item Holding out worlds or tasks costs little, but holding out tools does not. Detector rankings cross near 50\% tool coverage and reverse below it, because unseen tools map to inputs that training never reached, such as an untrained unknown-tool embedding. Dropout on tool inputs removes the GRU reversal and turns the other into a near tie. Training thresholds fail as well, and SE, a detector without untrained inputs, shows that a domain shift alone can move them (Section~\ref{sec:tools}).
\item No detector trained on AgentDrift ranks AgentDojo traces better than chance, not even one that is near perfect within AgentDojo, and within each benchmark surface content, such as generator templates or fixed attack strings, separates the classes. The released monitor, trained on neither, does transfer, but with its open backbone its released decision rule would still mostly raise false alerts at a 2\% attack rate (Section~\ref{sec:transfer}).
\end{itemize}
We turn these findings into seven recommendations (Section~\ref{sec:discussion}). We do not propose a detector. Our claims concern our own detectors, one reimplementation and one released monitor prompt, not the published detectors of Table~\ref{tab:audit} as released.

\begin{table*}[t]
\caption{Evaluation Protocols of the Audited Detector Studies}
\label{tab:audit}
\centering
{\scriptsize\renewcommand{\arraystretch}{1.1}\setlength{\tabcolsep}{4pt}
\begin{tabular}{@{}>{\raggedright\arraybackslash}p{2.0cm}>{\raggedright\arraybackslash}p{3.0cm}>{\raggedright\arraybackslash}p{2.1cm}>{\raggedright\arraybackslash}p{2.9cm}>{\raggedright\arraybackslash}p{2.9cm}>{\raggedright\arraybackslash}p{3.3cm}@{}}
\toprule
Work & Detector & Test attack share & Metrics & Shift tested & Cross-benchmark or within-target test \\
\midrule
TraceAegis \cite{traceaegis} & Hierarchical rules and LLM-summarized behavioral constraints & $\approx$50\% & P, R, F1, accuracy & None & No \\
Praetor \cite{praetor} & Automaton with embedding-based parameter guards & Classes scored separately & Attack success with intervals, benign task failure, latency & 20\% of tools replaced (benign traffic only) & Partial: benign traces from ToolBench, attacks from ASB \\
Agent-Sentry \cite{agentsentry} & XGBoost or rules, allowlist, then LLM judge & 38.7\% (Native-Trace 81.2\%) & Utility, block rate, FPR, FNR, F1 & Grouped by user task, prompt-style shift & Partial: Native-Trace with the same agents and tools, within-target control \\
AgentDrift \cite{agentdrift} & Surface-feature LR (benchmark check, not a detector) & 44.2\% & P, R, F1, AUROC, AUPRC & None (stratified split) & No \\
DriftNet \cite{driftnet} & Frozen step embeddings and a dual-head transformer & 45.0\% & P, R, F1, step F1, EM, IoU & Task-disjoint split & No \\
Trajectory Guard \cite{trajguard} & Siamese recurrent autoencoder & 45.5\% (external sets anomaly-only) & P, R, F1 (recall only on external sets) & External anomaly sources & Partial: external sets without benign traces \\
MCPShield \cite{mcpshield} & GNNs and classical models on content or tool metadata & 86.3\%, 49.7\%, 68.9\% & AUROC, AUPRC, F1, P, R, FPR over 3 seeds & Task-disjoint split & No (sources pooled) \\
Skynet \cite{skynet} & Graph contrastive encoder trained on benign traces & 16.5\% to 59.0\% & Accuracy, P, R, F1, FPR, ROC curves without area & Unseen anomaly types & Partial: classes drawn from different sources \\
TrajAD \cite{trajad} & Fine-tuned generative verifier & Balanced & P, R, macro-F1, joint exact match & Leave-one-domain-out (procedural errors, not attacks) & No \\
SentinelAgent \cite{sentinelagent} & Execution graph and LLM overseer & -- & Qualitative & None & No \\
LlamaFirewall \cite{llamafirewall} & Input classifier and few-shot LLM action auditor & -- (prompt level) & Attack success, utility, AUC, recall at 1\% FPR, FPR & Out-of-distribution jailbreak sets, prompt level & Partial: classifier trained on Meta's data, applied to AgentDojo \\
\bottomrule
\end{tabular}}
\par\vspace{2pt}\begin{minipage}{\textwidth}\footnotesize Attack or anomaly shares are our arithmetic from the class counts each paper reports, in the test sets that contain both classes. ``Partial'' means that the aspect is tested only in the limited form stated. P is precision, R recall, EM exact match and IoU intersection over union. Only Praetor evaluates adaptive attackers. The extended report gives the full protocol of each study, checked against the version audited.\end{minipage}
\end{table*}

\section{Related Work}
\label{sec:related}

\subsection{Detectors Over Agent Traces and How They Are Evaluated}

We identified the audited works through targeted searches of arXiv and Google Scholar up to September 2026. A work was included when it proposed or evaluated a trajectory-level or step-level detector of compromised or anomalous agent behavior on agent traces, and eleven works met this criterion. The audit is not a systematic review, since we followed no registered search protocol, and relevant detectors may have been missed. We left out MELON \cite{zhu2025melon}, which re-executes the live agent rather than scoring recorded traces, and guardrail benchmarks such as TraceSafe \cite{chen2026tracesafe} and ATBench \cite{li2026atbench}. The audited detectors range from rule, automaton and LLM-overseer monitors, such as TraceAegis, Praetor, Agent-Sentry and SentinelAgent, to learned sequence, graph and verifier models, such as DriftNet, Trajectory Guard, Skynet, MCPShield and TrajAD. LlamaFirewall pairs the PromptGuard 2 input classifier with AlignmentCheck, a few-shot LLM auditor that reasons over the whole execution trace without fine-tuning. AgentDrift \cite{agentdrift} is a benchmark rather than a detector, and its surface-feature logistic regression only checks that the benchmark cannot be solved from surface cues.

Table~\ref{tab:audit} summarizes each protocol, with every cell checked against the arXiv full text of the version audited. In the test sets that mix both classes, attacks or anomalies make up between 16.5\% and 86.3\% of the samples, by our arithmetic from the reported class counts. Four works report a threshold-free measure, an ROC or precision-recall summary or curve. The others report metrics at one operating point or no quantitative metric, and none varies the class ratio. Eight works test drift in a limited form, such as a task-disjoint split, unseen anomaly types or a prompt-style shift, and three not at all. Six works report no cross-benchmark or within-target test, and the other five report one only in part, for example with the same agents and tools or with anomalies but no benign traces. Only Praetor evaluates adaptive attackers.

\subsection{Agent-Security Benchmarks}

InjecAgent and AgentDojo measure how often injected tool outputs redirect an agent, and ASB \cite{zhang2025asb}, AgentDyn \cite{li2026agentdyn} and RAS-Eval \cite{fu2025raseval} extend the measurement to more scenarios and to dynamic or real tool execution. These benchmarks report attack success and utility for live agents and do not provide labeled corpora for training trace detectors, which AgentDrift does. Many defenses, MELON among them, act on the live agent or its inputs. The audited detectors and ours instead score recorded traces, which is closer to intrusion detection over logs.

\subsection{Base Rates and Distribution Shift}

Because precision depends on the class prior, precision and AUPRC values from test sets with different class ratios cannot be compared directly. Siblini et al. \cite{siblini2020} calibrate average precision to a reference prior in closed form, and Boyd et al. \cite{boyd2013} compare estimators of the area under the precision-recall curve. McDermott et al. \cite{mcdermott2024} question the general preference for AUPRC under imbalance, so we report AUROC next to it. Sommer and Paxson \cite{sommer2010} place anomaly-based intrusion detection outside the closed-world assumption of most machine learning, Arp et al. \cite{arp2022} catalogue pitfalls such as sampling bias and spurious correlations, TESSERACT \cite{pendlebury2019} shows how biased splits inflate malware classification results, and conformal evaluation rejects decisions on drifting samples \cite{barbero2022}. These works treat drift mainly as a function of time. Out-of-distribution accuracy is often almost linearly correlated with in-distribution accuracy \cite{miller2021}, but Teney et al. \cite{teney2023} show that the correlation can be inverted, so that selecting by in-distribution accuracy selects against robustness. Our results give a concrete, measurable instance of such an inversion.

\section{Study Design}
\label{sec:design}

\subsection{Threat and Deployment Model}

We consider an indirect injector who controls the content of tool outputs that the agent reads, has no access to the agent's model or to the detector, and does not adapt to the detector. The goal is to make the agent act on the attacker's behalf, for example by sending data or money, or, in denial-of-service (DoS) variants, to stop it from completing its task. The defender scores the recorded trace of each finished session once, and an alert sends the session to an analyst (Fig.~\ref{fig:overview}a). Our setting is therefore a post hoc audit. Recall counts hijacked sessions identified afterwards rather than prevented actions, and alerts are counted per 1,000 audited sessions. The prevalence grid of 1\% to 10\% stands for the share of audited sessions that contain a successful hijack, and its lower end is set by our pool sizes.

The content-blind detectors read at most tool names, argument counts and observation lengths. SE and the content reference also read argument values and observation text, including the injected text, and the monitor reads tool names and arguments but no observation. The attacker directly writes the observation length, since the injected observation is one of the observations. The other behavioral features depend on the actions the hijack induces, which the attacker influences through the injected goal but does not set.

\subsection{Benchmarks}

\emph{AgentDrift.} AgentDrift \cite{agentdrift} release 1.0 contains 12,536 synthetic trajectories from five domains (email, banking, web, coding and medical), generated by a quantized Llama-3.3-70B-Instruct with prompts and scripts that are not released. Its 103 tools are disjoint across domains. The negatives are 4,000 benign trajectories and 1,500 hard negatives, and 1,500 failed attacks serve as training negatives and are scored separately at test. The 5,536 attacked trajectories are full, partial or delayed hijacks. Every trajectory carries a world record (user, e-mail address, company, contacts and date). We use the released stratified split (8,775 training, 1,880 validation and 1,881 test trajectories) and the released task-disjoint split. Banking has 256 full hijacks against 720 in each other domain, which confounds comparisons across domains that involve banking.

\emph{AgentDojo.} We use the recorded runs without defenses of \texttt{gpt-4o-2024-05-13}, \texttt{claude-3-5-sonnet-20241022}, \texttt{gemini-1.5-pro-002}, \texttt{gpt-4-turbo-2024-04-09} and \texttt{gemini-2.0-flash-001} over the workspace, travel, banking and Slack suites of AgentDojo \cite{agentdojo}. A run without an attack is benign, and a run with an attack is compromised when the benchmark's security check reports that the attacker's goal was reached and unsuccessful otherwise. After runs with fewer than two tool calls are removed, population P0 holds 2,294 compromised and 392 benign traces, and the 4,987 unsuccessful attacks are scored separately. Of the 67 AgentDojo tools, 7 occur in AgentDrift, and 69.8\% of P0 traces contain no tool seen in AgentDrift. Two properties act as confounds. \texttt{gpt-4o-2024-05-13} contributes 81.3\% of the compromised traces, and 101 compromised traces come from DoS variants, which AgentDojo labels compromised exactly when the user task was not completed. Appendix~\ref{app:dojo} describes the extraction and the populations.

\subsection{Detectors}

\emph{Content-blind detectors.} Representation A counts how often each training tool is called, so a trajectory with no known tool maps to the zero vector. Representation B has ten behavioral features that do not name tools, such as trajectory length, tool diversity, the share of write-like calls and the maximum and mean observation length (Appendix~\ref{app:detectors}). Logistic regression on B gives LR-B, and on A and B together LR-C. The gated recurrent unit (GRU) \cite{cho2014gru} reads the sequence of tool tokens, with one token for unknown tools, together with the ten features of B. No AgentDrift training trajectory contains an unknown tool, so the embedding of that token is never trained. Two dropout variants train the missing input. LR-C-drop adds a copy of the training set whose tool block is zeroed, which is equivalent to dropping the tool block with probability 0.5, and GRU-drop0.2 replaces each training token by the unknown token with probability 0.2.

\emph{Step embeddings.} SE is our reimplementation of the DriftNet architecture \cite{driftnet}, whose code and weights have not been released, so none of its numbers is DriftNet's own. Each step is written as the tool name, the sorted arguments and the observation, embedded with a frozen sentence encoder, and the sequence is scored by a small transformer trained on AgentDrift. A held-out tool name is then an ordinary point in a space that the encoder already covers, so SE has no untrained input.

\emph{Content reference.} TF-IDF is a class-weighted logistic regression on term frequency-inverse document frequency features \cite{salton1988tfidf} of the full text of every step, including argument values and the observation text, and thus the injected text.

\emph{Released action monitor.} AlignmentCheck \cite{llamafirewall} shows an LLM the user's request and the agent's actions so far and asks whether the latest action pursues a goal that the user did not set. We use its released prompt and output schema verbatim and scan each step against the trace before it, as the released scanner does. Its native decision, the released decision rule, flags a trace when any step is flagged. Each step shows the tool name and the arguments but no tool observation, so the monitor never reads the injected text. MON-L runs the prompt on Llama 3.3 70B Instruct \cite{llama3} in a 4-bit quantization \cite{gptq}, served with vLLM \cite{vllm} at temperature 0. Its graded step score is the probability of \texttt{true} at the conclusion token, and its trace score is the maximum over steps. MON-A runs the prompt on Claude Sonnet 5.5, and its coarse trace score is the number of flagged steps. The monitor is trained on neither benchmark. It is not LlamaFirewall as deployed, since there is no PromptGuard layer, the scan runs after the fact and neither backbone is Meta's default. AgentDrift was generated by a Llama 3.3 70B model, so MON-L may be familiar with its traces, and the monitor's main claims therefore rest on AgentDojo.

\subsection{Settings and Tool Coverage}
\label{sec:settings}

A trajectory is a sequence of tool-call steps
\begin{equation}
\tau = (s_1,\ldots,s_m), \qquad s_t = (a_t, u_t, o_t),
\end{equation}
where $a_t$ is the tool name, $u_t$ the arguments and $o_t$ the observation returned by the tool. The label $y(\tau)$ is 1 for an attacked AgentDrift trajectory or a compromised AgentDojo trace. A detector assigns a score $f(\tau)$ and flags the trajectory when the score reaches a threshold $\theta$,
\begin{equation}
\hat{y}_\theta(\tau) = \mathbf{1}\left[f(\tau)\ge\theta\right].
\end{equation}
In training, the failed attacks and hard negatives of AgentDrift count as negatives. For AgentDrift, evaluation uses pool E, which holds the benign, hard-negative and attacked test trajectories, and for AgentDojo it uses P0. With $\mathcal{V}$ the tool vocabulary of the benign training trajectories, the tool coverage of a test set $\mathcal{D}$ is the share of its tool calls whose tool appears in $\mathcal{V}$,
\begin{equation}
\kappa(\mathcal{D};\mathcal{V}) = \frac{\sum_{\tau\in\mathcal{D}}\sum_{t=1}^{m(\tau)}\mathbf{1}\left[a_t\in\mathcal{V}\right]}{\sum_{\tau\in\mathcal{D}}m(\tau)}.
\end{equation}

Table~\ref{tab:settings} defines the settings. R1 is the released stratified split. R1-wd re-splits the pooled data over five seeds so that no generation world is shared, and R2 is the released task-disjoint split. R3 holds out one domain, and with it all its tools, so $\kappa=0$. Inside R1, the graded-coverage control masks a random share of the training tool names at test time, which lowers coverage without changing the domain. R4 trains on AgentDrift and tests on AgentDojo, changing the ecosystem, the trace generator, attack construction and label semantics at once. The within-AgentDojo control trains and tests on P0 by five-fold cross-validation grouped by suite and user task (UT), repeated five times.

\begin{table*}[t]
\caption{Evaluation Settings With Their Tool Coverage}
\label{tab:settings}
\centering
{\footnotesize\renewcommand{\arraystretch}{1.08}\setlength{\tabcolsep}{4pt}
\begin{tabular}{@{}>{\raggedright\arraybackslash}p{2.2cm}>{\raggedright\arraybackslash}p{5.0cm}>{\raggedright\arraybackslash}p{4.6cm}>{\raggedright\arraybackslash}p{1.1cm}>{\raggedright\arraybackslash}p{3.7cm}@{}}
\toprule
Setting & Training data & Test pool ($n_-/n_+$) & $\kappa$ & Shift introduced \\
\midrule
R1 & Stratified training split (8,775) & Stratified test split, 825/831 & 1.000 & None (reference) \\
R1-wd & Pooled corpus with no shared world, split seeds 0 to 4 (8,711 to 8,764) & 786 to 811 / 863 to 924 & 1.000 & World identity \\
R2 & Task-disjoint training split (9,081) & 35 unseen tasks, 729/775 & 1.000 & Task \\
R3 & Four domains (8,446, or 8,840 with banking held out) & Held-out domain, 935/1,020 (banking 935/626) & 0.000 & Domain and all its tools \\
R4 & AgentDrift training and test splits pooled (10,656) & AgentDojo P0, 392/2,294 (P0F adds 4,987 unsuccessful attacks) & 0.075 & Benchmark, ecosystem, generator, labels \\
Within AgentDojo, UT & Four fifths of P0, 5 repetitions & Held-out folds of P0 & 0.976 & Unseen user tasks \\
\bottomrule
\end{tabular}}
\par\vspace{2pt}\begin{minipage}{\textwidth}\footnotesize $\kappa$ is the share of test tool calls whose tool occurs in the benign training trajectories, from (3). The graded-coverage control of Fig.~\ref{fig:coverage} lowers $\kappa$ inside R1 by masking tool names at test time. Within-AgentDojo rows are cross-validated on P0, so their test pool is P0 itself.\end{minipage}
\end{table*}

\subsection{Metrics, Thresholds and Statistics}

Prevalence shift leaves the class-conditional score distributions unchanged, so precision at prevalence $\pi$ follows from the true- and false-positive rates at a threshold,
\begin{equation}
\mathrm{Prec}(\theta;\pi) = \frac{\pi\,\mathrm{TPR}(\theta)}{\pi\,\mathrm{TPR}(\theta) + (1-\pi)\,\mathrm{FPR}(\theta)}.
\label{eq:prec}
\end{equation}
Our ranking measure at low prevalence is the prevalence-weighted average precision over the thresholds $\theta_k$ of the full test pool,
\begin{equation}
\mathrm{wAUPRC}@\pi = \sum_k \big(\mathrm{TPR}(\theta_k) - \mathrm{TPR}(\theta_{k-1})\big)\,\mathrm{Prec}(\theta_k;\pi),
\end{equation}
which is the calibrated average precision of Siblini et al. \cite{siblini2020}. It uses every positive at every prevalence, so its intervals come from a bootstrap of the full pool. We report $\pi$ of 1\%, 2\%, 5\% and 10\%.

Operating points use thresholds fixed without test data. The main one,
\begin{equation}
\theta_{r80} = Q_{0.20}\big(\{f(\tau) \mid \tau\in\mathcal{D}_{\mathrm{train}},\ y(\tau)=1\}\big),
\end{equation}
is the 20th percentile of the training-positive scores, so it flags 80\% of training positives. For the monitor we also report its native decision. In the within-AgentDojo control, thresholds are recomputed on each training fold.

Intervals are 95\% percentile intervals from 2,000 bootstrap replicates \cite{efron1993bootstrap}, stratified by class for AgentDrift. AgentDojo intervals resample the 117 suite and user-task clusters \cite{field2007bootstrap}. Where a detector has several fits, such as GRU training seeds, each replicate draws one fit. Paired differences resample the same rows for both detectors. We call a difference distinguishable from zero when its interval excludes zero, and we apply the Holm correction \cite{holm1979} within pre-defined families, namely the five R3 folds, the comparisons at one prevalence and the detectors of one cross-benchmark configuration. A result whose decision depends on the resampling unit or on the correction is called borderline. Because an R3 AUROC ranks trajectories within one domain, each R3 fold is compared with R1dm, the R1 model scored on the R1 test trajectories of the held-out domain. A ranking reversal needs an in-distribution advantage in R1dm and an opposite R3 difference, both distinguishable after Holm correction (Appendix~\ref{app:rules}).

\emph{Pre-specification.} The analyses follow pre-specified internal plans, none of them externally registered, and only parts of them were prospective. The cross-benchmark plan was completed after the transfer results were known, and a revision plan was written after an independent review had seen exploratory versions of most analyses, so most results here are confirmatory re-runs rather than prospective tests. The monitor is the exception. Its plan was registered by hash before any monitor call, and its predictions were prospective. Appendix~\ref{app:plans} lists what was known when each plan was written, the readings added afterwards, and the monitor's predictions with their outcomes.

\section{Rare Attacks: A High AUROC, Mostly False Alerts}
\label{sec:rare}

\emph{Finding 1. A high AUROC does not imply a usable operating point once attacks are rare, and the precision that a detector reaches depends as much on the mix of negatives as on prevalence.}

\begin{table*}[t]
\caption{Detection in R1 at Controlled Attack Prevalence}
\label{tab:prevalence}
\centering
{\footnotesize\renewcommand{\arraystretch}{1.12}\setlength{\tabcolsep}{3.3pt}
\begin{tabular}{@{}lcccccccc@{}}
\toprule
 & & \multicolumn{3}{c}{wAUPRC at prevalence} & Precision & & & Alerts per 1,000 \\
\cmidrule(lr){3-5}
Detector & AUROC & 1\% & 2\% & 10\% & at 2\% & Recall & FPR & at 2\% \\
\midrule
LR-B (behavioral) & \cit{0.830}{0.810}{0.849} & \cit{0.098}{0.076}{0.175} & \cit{0.152}{0.119}{0.231} & \cit{0.403}{0.354}{0.462} & \cit{0.054}{0.048}{0.061} & $0.795$ & $0.284$ & 294 \\
LR-C (combined) & \cit{0.879}{0.862}{0.895} & \cit{0.156}{0.121}{0.261} & \cit{0.232}{0.183}{0.330} & \cit{0.523}{0.466}{0.590} & \cit{0.076}{0.067}{0.088} & $0.771$ & $0.192$ & 203 \\
GRU (sequence) & \cit{0.936}{0.922}{0.947} & \cit{0.265}{0.177}{0.442} & \cit{0.382}{0.282}{0.537} & \cit{0.699}{0.628}{0.770} & \cit{0.238}{0.184}{0.328} & $0.682$ & $0.045$ & 58 \\
TF-IDF (content) & \cit{0.999}{0.999}{1.000} & \cit{0.975}{0.960}{0.991} & \cit{0.984}{0.973}{0.994} & \cit{0.996}{0.992}{0.998} & 1.000 ($\geq$0.771) & $0.763$ & 0.000 ($\leq$0.004) & 15 \\
\midrule
MON-L at $\theta_{r80}$ & \cit{0.856}{0.837}{0.874} & \cit{0.048}{0.040}{0.067} & \cit{0.089}{0.075}{0.117} & \cit{0.338}{0.298}{0.395} & \cit{0.083}{0.072}{0.096} & $0.769$ & $0.173$ & 185 \\
MON-L, native decision & as above &  &  &  & \cit{0.073}{0.064}{0.082} & $0.847$ & $0.221$ & 233 \\
MON-A, native decision & \cit{0.817}{0.796}{0.837}$^\ddagger$ & \cit{0.038}{0.032}{0.060}$^\ddagger$ & \cit{0.072}{0.062}{0.100}$^\ddagger$ & \cit{0.287}{0.256}{0.333}$^\ddagger$ & \cit{0.054}{0.049}{0.060} & $0.874$ & $0.313$ & 324 \\
Random scorer & $0.500$ & \cit{0.011}{0.009}{0.016} & \cit{0.022}{0.019}{0.027} & \cit{0.102}{0.093}{0.114} & $0.020$ & -- & -- & -- \\
\bottomrule
\end{tabular}}
\par\vspace{2pt}\begin{minipage}{\textwidth}\footnotesize Pool E of R1 has 831 attacked and 825 negative trajectories (600 benign and 225 hard negatives). Precision, recall, FPR and alerts are at $\theta_{r80}$, except in the native-decision rows, which use the released decision rule. Precision and alerts per 1,000 sessions follow (\ref{eq:prec}) under the pool-E negative mix. TF-IDF flags no negative, so its FPR is the exact 95\% upper bound and its precision the lower bound implied by it. The random-scorer row gives the mean and the 2.5th to 97.5th percentiles over 2,000 random scorers. $^\ddagger$MON-A ranks traces by the number of flagged steps, so its AUROC and wAUPRC are coarse.\end{minipage}
\end{table*}

\begin{figure*}[t]
\centering
\includegraphics[width=\textwidth]{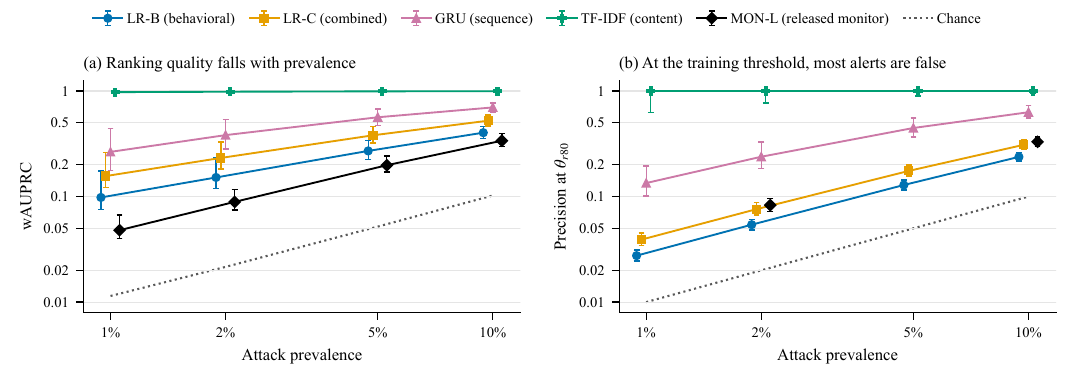}
\caption{Detection quality in R1 as attack prevalence falls. (a) Prevalence-weighted AUPRC. (b) Precision at the training threshold $\theta_{r80}$. The bar of TF-IDF extends to the lower bound implied by its zero false positives, and MON-L is shown at the two prevalences for which its precision was computed. Both axes are logarithmic, bars are 95\% bootstrap intervals, and the dotted line is chance.}
\label{fig:prevalence}
\end{figure*}

Table~\ref{tab:prevalence} and Fig.~\ref{fig:prevalence} show detection in R1 as attacks become rarer. LR-C has an AUROC of 0.879, yet at 2\% prevalence its wAUPRC is 0.232 and its precision at $\theta_{r80}$ is 0.076, with 12.2 false alerts for every true one. The GRU ranks better, with an AUROC of 0.936 and a precision of 0.238 at 2\%, but its test recall at $\theta_{r80}$ is 0.682, short of the 0.80 that its threshold targets. Below the floor set by our pool size, (\ref{eq:prec}) gives LR-C a precision of 0.0040 at 0.1\% prevalence. The estimator matters as well. An earlier version of this study subsampled the positives to each target prevalence, and its interval for LR-C at 2\% was 2.46 times wider than the weighted one.

The negative mix matters as much as prevalence. At $\theta_{r80}$, LR-C flags 0.118 of benign trajectories, 0.387 of hard negatives and 0.218 of failed attacks. Its precision at 2\% is therefore 0.117 [0.097, 0.145] against benign negatives only, 0.076 with the 27\% hard negatives of pool E, and 0.039 [0.034, 0.047] against hard negatives only. LR-B, LR-C and the GRU all flag hard negatives more often than benign trajectories, so a precision figure means little unless the negatives behind it are stated.

Two detectors are near perfect in distribution. TF-IDF reaches an AUROC of 0.999 and flags no negative of pool E at $\theta_{r80}$, and SE reaches 0.998 with an FPR of 0.001. Section~\ref{sec:content} shows that this reflects how AgentDrift was generated rather than a solved task. The released monitor, which judges actions without reading observations, reaches an AUROC of 0.856 on R1, slightly below LR-C (difference $-$0.023 [$-$0.043, $-$0.002]). Its native decision flags 0.847 of attacked and 0.221 of negative trajectories, for an implied precision of 0.073 at 2\% prevalence, and MON-A's decision reaches 0.054. Of the detectors in Table~\ref{tab:prevalence}, only TF-IDF keeps a precision above 0.5 at a 2\% attack rate.

\section{Unseen Tools: Rankings Reverse, Thresholds Fail}
\label{sec:tools}

\emph{Finding 2. Holding out worlds or tasks costs little, but removing known tools reverses detector rankings, gradually as coverage falls. On AgentDrift the reversal comes mainly from inputs that were never trained, and dropout on tool inputs removes the GRU reversal and turns the LR-C reversal into a near tie. Training thresholds fail as well, and that failure also occurs without untrained inputs.}

\subsection{Worlds and Tasks Cost Little}

R1 reuses generation worlds, since for 99.3\% of its test trajectories the world also occurs in training. Holding worlds out nevertheless shows little inflation. Over five split seeds of R1-wd, no decrease in AUROC is distinguishable from zero in any of 25 detector and seed pairs. On unseen tasks (R2) no change is distinguishable in either direction. With task-cluster resampling, R2 $-$ R1 AUROC is +0.023 [$-$0.005, +0.049] for LR-C and $-$0.020 [$-$0.054, +0.008] for the GRU. Unlike MCPShield's tool-name metadata, which falls from 0.898 to 0.640 under task-disjoint splits \cite{mcpshield}, a logistic regression on our tool counts alone scores 0.788 in R2 against 0.777 in R1. All tasks of an AgentDrift domain share one tool set, so both splits keep full coverage (Table~\ref{tab:settings}).

\subsection{Removing Known Tools Gradually}

\begin{figure*}[t]
\centering
\includegraphics[width=\textwidth]{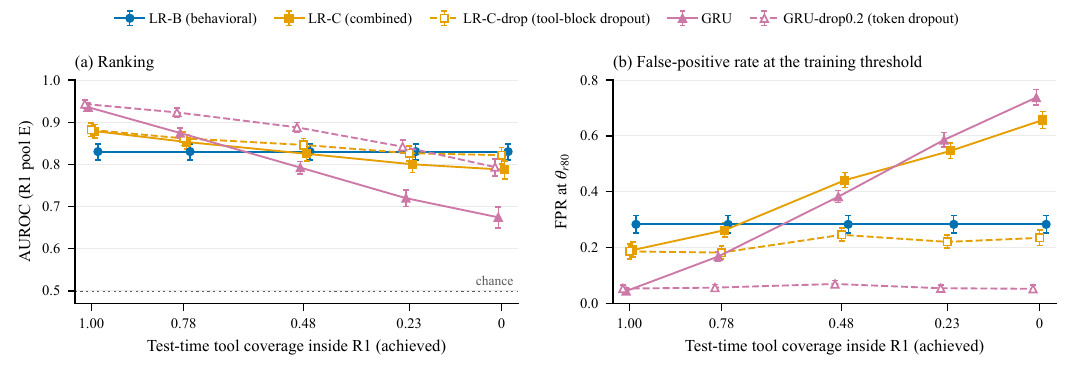}
\caption{Removing tool identity at test time inside R1. (a) AUROC and (b) FPR at $\theta_{r80}$ against the achieved tool coverage. Dashed lines with hollow markers are the dropout variants, and bars are 95\% intervals conditional on the drawn masks and seeds. LR-B uses no tool feature and is constant.}
\label{fig:coverage}
\end{figure*}

To separate the loss of known tools from the change of domain, we mask a random share of the 103 training tool names at test time inside R1 (Fig.~\ref{fig:coverage}). At achieved coverages of 1.000, 0.775, 0.483, 0.225 and 0.000, LR-C falls from 0.879 through 0.853, 0.825 and 0.800 to 0.787, and the GRU from 0.936 to 0.674, while LR-B, which uses no tool feature, stays at 0.830. LR-C $-$ LR-B is +0.049 [+0.034, +0.064] at full coverage, $-$0.005 [$-$0.016, +0.006] at 0.483 and $-$0.043 [$-$0.055, $-$0.031] at zero, and GRU $-$ LR-C goes from +0.057 [+0.043, +0.071] to $-$0.113 [$-$0.129, $-$0.096]. For both pairs the crossing lies between coverages of 0.775 and 0.483, so a deployment in which about half the tools are new already removes the in-distribution advantage, well before the disjoint-vocabulary case of R3. The cross-benchmark setting of Section~\ref{sec:transfer} sits near the lower end of this range, at $\kappa$ = 0.075. Thresholds degrade over the same range. The FPR of LR-C at $\theta_{r80}$ rises from 0.192 to 0.657, and that of the GRU from 0.045 to 0.737. Dropout keeps the FPR nearly flat, between 0.053 and 0.070 for GRU-drop0.2, and slows the ranking loss, since GRU-drop0.2 falls from 0.943 to 0.794. The released monitor also relies on what tool names mean. Masking every name lowers MON-L's AUROC on R1 by 0.091, to 0.765 [0.743, 0.788].

\subsection{Holding Out a Domain}

Holding out a domain removes all its tools. Table~\ref{tab:domain} compares each R3 fold with R1dm, the R1 model scored on the same domain's R1 test trajectories. LR-B changes in no fold. LR-C loses 0.065 to 0.155 AUROC in web, email and coding, and the GRU falls in every fold, to between 0.639 and 0.736. The in-distribution orderings do not survive. After Holm correction, LR-C ranks below LR-B in all five held-out domains and the GRU below LR-C in all five, and the reversal rule, which also requires a distinguishable in-distribution advantage, finds a C-versus-B reversal in three domains and a GRU-versus-C reversal in four. The pre-specified analysis plan for the GRU did not anticipate the GRU-versus-C reversal, so we treat that reversal as exploratory. The detectors without untrained inputs keep their places. TF-IDF stays between 0.980 and 0.996, SE between 0.904 and 0.959 and above LR-B in every fold, distinguishably in four, and MON-L between 0.829 and 0.925, above LR-C and the GRU in every fold.

\begin{table}[t]
\caption{AUROC per Held-Out Domain}
\label{tab:domain}
\centering
{\footnotesize\renewcommand{\arraystretch}{1.08}\setlength{\tabcolsep}{4pt}
\begin{tabular}{@{}lccccc@{}}
\toprule
 & Medical & Banking & Web & Email & Coding \\
\midrule
\multicolumn{6}{@{}l}{\emph{Same domain inside R1 (R1dm)}} \\
LR-B & $0.793$ & $0.832$ & $0.886$ & $0.804$ & $0.845$ \\
LR-C & $0.786$ & $0.853$ & $0.927$ & $0.894$ & $0.894$ \\
GRU & $0.916$ & $0.915$ & $0.925$ & $0.966$ & $0.944$ \\
\midrule
\multicolumn{6}{@{}l}{\emph{Domain and all its tools held out (R3)}} \\
LR-B & $0.808$ & $0.854$ & $0.896$ & $0.792$ & $0.852$ \\
LR-C & $0.745$ & $0.828$ & $0.862^{\downarrow\dagger}$ & $0.739^{\downarrow\dagger}$ & $0.814^{\downarrow\dagger}$ \\
LR-C-drop & $0.797$ & $0.853$ & $0.894^{\S}$ & $0.777^{\S}$ & $0.847^{\S}$ \\
GRU & $0.639^{\downarrow\dagger}$ & $0.736^{\downarrow\dagger}$ & $0.726^{\downarrow}$ & $0.671^{\downarrow\dagger}$ & $0.695^{\downarrow\dagger}$ \\
GRU-drop0.2 & $0.747$ & $0.848$ & $0.870$ & $0.766$ & $0.811$ \\
SE & $0.959$ & $0.917$ & $0.904$ & $0.950$ & $0.938$ \\
TF-IDF & $0.996$ & $0.989$ & $0.980$ & $0.990$ & $0.987$ \\
MON-L & $0.829$ & $0.864$ & $0.925$ & $0.890$ & $0.855$ \\
\bottomrule
\end{tabular}}
\par\vspace{3pt}\begin{minipage}{\columnwidth}\footnotesize LR-B is the behavioral and LR-C the combined logistic regression, GRU the sequence model, SE the step-embedding detector, TF-IDF the content reference and MON-L the released monitor. $^{\downarrow}$R3 decrease against R1dm distinguishable from zero after Holm correction over the five domains. $^{\dagger}$Reversal by the rule of Appendix~\ref{app:rules}, of LR-C against LR-B or of the GRU against LR-C. $^{\S}$LR-C-drop meets the rule against LR-B, but the two rank trajectories almost identically (Spearman correlation of at least 0.997). R1dm pools hold 111 to 180 positives and 165 negatives per domain, R3 pools 626 to 1,020 positives and 935 negatives. Comparisons across folds that involve banking are confounded by its hijack mix.\end{minipage}
\end{table}

\subsection{Why the Rankings Reverse}

The reversals come from inputs that training never reached. At $\kappa=0$ the tool block of LR-C is an all-zero raw vector, which the training scaler maps to a constant $c$. Writing the logit as $g(\tau)=w_A^\top z_A(\tau)+w_B^\top z_B(\tau)+b$, with standardized tool and behavioral blocks $z_A$ and $z_B$, every held-out trajectory receives
\begin{equation}
g(\tau) = w_B^\top z_B(\tau) + \big(w_A^\top c + b\big).
\end{equation}
The ranking within a fold therefore rests on LR-C's behavioral weights alone, which were fitted together with the tool features and rank held-out trajectories worse than LR-B does. The offset $w_A^\top c$ is +1.653 to +1.899 logits, a tool contribution that only 10.7\% to 15.0\% of training trajectories reach. It pushes every held-out trajectory toward the attack side, so recall and FPR rise together. The GRU shows the same effect more strongly. Weight decay shrinks the norm of its untrained unknown-token embedding well below those of the trained tool embeddings, and through the tool pathway held-out negatives receive a mean logit 6.523 higher than training negatives, against 0.476 in R1.

Dropout trains the missing input and shrinks both offsets. GRU-drop0.2 meets the reversal rule in no fold, at no in-distribution cost (R1 AUROC 0.943 against 0.936). LR-C-drop still meets the rule against LR-B in three folds, but the two rank trajectories almost identically (Spearman correlation of at least 0.997), so we read these as near ties.

\subsection{Thresholds Fail for a Wider Reason}

Training thresholds fail in a detector-specific way. LR-C over-alerts in every held-out domain, with an FPR at $\theta_{r80}$ that rises from between 0.121 and 0.321 in R1dm to between 0.503 and 0.722 in R3. With its training seeds pooled, the GRU over-alerts in three folds after Holm correction, with an R3 FPR of 0.438 to 0.764 against at most 0.071 in R1dm. The threshold of each single seed fails in every fold, and the seeds disagree on the direction. LR-B's threshold carries over in four of five domains. SE shows that this failure does not need an untrained input. On its full logit, held-out negatives sit 5.161, 5.795 and 7.420 logits above training negatives in medical, coding and web, offsets as large as the GRU's, and its FPR at $\theta_{r80}$ rises from 0.001 in R1 to 0.046, 0.049 and 0.106 in those domains, while in banking and email its recall falls to 0.345 and 0.620. A domain shift in a trained input space can thus produce logit offsets as large as those of an untrained input, and with them threshold failure. Recalibration without labels helps only in part. A threshold set on 200 unlabeled held-out trajectories to reproduce the R1 alert rate brings LR-C's fold-mean FPR down to 0.217 to 0.312, but under a post hoc reading at 2\% prevalence the GRU's recall collapses to 0.123 to 0.273.

\subsection{An Attacker-Written Cue}

What remains of the behavioral signal at zero coverage is largely an observation-length cue. The mean observation length carries the largest standardized weight in LR-B ($-$1.861 [$-$1.965, $-$1.770]), and even benign-labeled steps of attacked trajectories are short, which points to the generator. Without the two observation-length features LR-B loses 0.112 to 0.221 AUROC in every R3 fold, while the two features alone come within 0.053 of the full model. The attacker writes this quantity. Appending 1,200 characters to the injected observation of each attacked R1 trajectory lowers the recall at $\theta_{r80}$ of LR-C from 0.771 to 0.033, of the GRU from 0.682 to 0.000 and of SE from 0.796 to 0.531. Within AgentDojo the cue points the other way, since the mean observation length ranks compromised traces slightly above benign ones (AUROC 0.558), and padding slightly raises LR-C's recall. We tested only the effect on scores, not whether padded injections would still succeed.

\section{A New Benchmark: Trained Detectors Do Not Transfer}
\label{sec:transfer}

\emph{Finding 3. No detector trained on AgentDrift ranks AgentDojo traces better than chance, and within each benchmark surface content separates the classes. A released monitor that is trained on neither benchmark transfers, but its native decision does not solve the rare-attack problem of Section~\ref{sec:rare}.}

\begin{table*}[t]
\caption{Cross-Benchmark Transfer to AgentDojo and the Within-AgentDojo Control}
\label{tab:transfer}
\centering
{\footnotesize\renewcommand{\arraystretch}{1.12}\setlength{\tabcolsep}{4pt}
\begin{tabular}{@{}lcccc>{\raggedright\arraybackslash}p{3.8cm}@{}}
\toprule
 & Transfer AUROC & Within AgentDojo & Within $-$ transfer & Transfer AUROC, & Pre-specified reading \\
Detector & on P0 & AUROC (UT) & (paired) & P0F & (Appendix~\ref{app:rules}) \\
\midrule
LR-B (behavioral) & \cit{0.482}{0.436}{0.521} & \cit{0.741}{0.688}{0.792} & \cit{+0.263}{+0.198}{+0.323} & \cit{0.626}{0.575}{0.669} & Weak representation \\
LR-C (combined) & \cit{0.473}{0.427}{0.512} & \cit{0.797}{0.747}{0.842} & \cit{+0.327}{+0.268}{+0.382} & \cit{0.626}{0.574}{0.670} & Mixed, reasonable on C1 only \\
GRU (sequence) & \cit{0.513}{0.464}{0.551} & \cit{0.840}{0.784}{0.884} & \cit{+0.329}{+0.264}{+0.388} & \cit{0.645}{0.593}{0.689} & Transfer failure (UT), IT not reasonable \\
SE (step embeddings) & \cit{0.481}{0.399}{0.560} & \cit{0.956}{0.916}{0.984} & \cit{+0.478}{+0.389}{+0.563} & \cit{0.572}{0.490}{0.657} & Transfer failure (UT), IT not run \\
TF-IDF (content) & \cit{0.520}{0.473}{0.565} & \cit{0.996}{0.992}{0.999} & \cit{+0.477}{+0.430}{+0.523} & \cit{0.626}{0.578}{0.666} & Transfer failure (UT and IT) \\
\midrule
MON-L (released monitor) & \cit{0.930}{0.909}{0.951} & -- & -- & \cit{0.916}{0.897}{0.932} & Not trained, transfers \\
MON-A (released monitor) & \cit{0.934}{0.915}{0.952} & -- & -- & \cit{0.918}{0.899}{0.936} & Not trained, transfers \\
\bottomrule
\end{tabular}}
\par\vspace{2pt}\begin{minipage}{\textwidth}\footnotesize Transfer models are trained on AgentDrift, and within models by cross-validation grouped by user task (UT). Intervals resample suite and user-task clusters. Every within-minus-transfer gap is distinguishable from zero after Holm correction over the representations. P0F adds the 4,987 unsuccessful attacks to P0 as negatives, and C1 removes the 65 benign traces whose user task was an injection task. IT groups compromised traces by injection goal instead (Appendix~\ref{app:rules}). The monitor is not trained, so the within-AgentDojo columns do not apply to it.\end{minipage}
\end{table*}

\subsection{Trained Detectors Lie Near Chance}
\label{sec:notransfer}

\begin{figure}[t]
\centering
\includegraphics[width=\columnwidth]{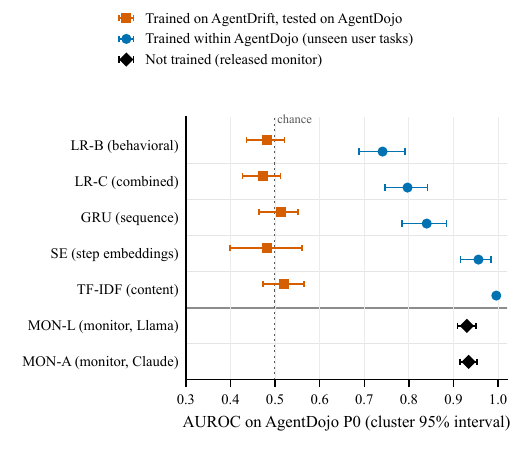}
\caption{AUROC on AgentDojo P0. Squares are detectors trained on AgentDrift, circles the same representations trained within AgentDojo on other user tasks, and diamonds the released monitor, which is not trained. Bars are 95\% cluster-bootstrap intervals, and the dotted line is chance.}
\label{fig:transfer}
\end{figure}

Table~\ref{tab:transfer} and Fig.~\ref{fig:transfer} report AUROC on AgentDojo P0. Every detector trained on AgentDrift lies between 0.473 and 0.520, with an interval that includes 0.5. At its training threshold, transferred LR-C flags 0.224 of benign and 0.225 of compromised traces. Composition does not explain the failure. With positives reweighted to the benign model mix, LR-C transfers at 0.486 [0.435, 0.530].

The within-AgentDojo control shows that the target can be learned. Trained inside AgentDojo on other user tasks, the same representations reach 0.741 to 0.996, and every within-minus-transfer gap is distinguishable from zero after Holm correction. The pre-specified rule (Appendix~\ref{app:rules}) reads this as a failure of transfer for the GRU, SE and TF-IDF. For LR-B the within-AgentDojo detection does not reach the rule's bar either, so its features are weak on the target as well, and LR-C's outcome is mixed, since it misses the bar on P0 by 0.003 AUROC but meets it on a population without injection-task benign runs. The population matters in a further way. With the unsuccessful attacks added as negatives (P0F), transfer rises to 0.626 for LR-C and 0.645 for the GRU, because the AgentDrift-trained models score unsuccessful attacks below benign runs.

\subsection{Both Benchmarks Are Separable by Surface Content}
\label{sec:content}

The content reference explains how a detector can be near perfect on one benchmark and useless on the other. On AgentDrift every trajectory category carries its own generator template. The token ``2026'', which has the strongest negative weight, occurs in 0.742 of benign and 0.848 of failed-attack training trajectories, in none of the hard negatives and in 0.046 of attacked ones. Masking world strings, anonymizing names or removing digits changes the R1 AUROC by at most 0.001, so the separation is spread over many template tokens. Within AgentDojo, TF-IDF largely detects that an injection is present. Five fixed attack strings occur in 0.996 of compromised and 0.947 of unsuccessful-attack traces and in no benign trace, and used as a fixed rule they give an AUROC of 0.998. Content does not rescue transfer, since TF-IDF transfers at 0.520 and flags no AgentDojo trace at its training threshold. A detector that scores well on either benchmark may therefore have learned how the data were generated rather than how a hijacked agent behaves.

\subsection{A Released Monitor Transfers but Still Over-Alerts}

AlignmentCheck, trained on neither benchmark, does transfer. MON-L reaches an AUROC of 0.930 [0.909, 0.951] on P0, and 0.940 [0.928, 0.952] with positives reweighted to the benign model mix, and it exceeds every AgentDrift-trained detector on the same traces by +0.410 to +0.457. MON-A reaches 0.934 [0.915, 0.952]. This suggests that the failure of Section~\ref{sec:notransfer} lies in training on one benchmark rather than in the task. Meta evaluated the prompt on AgentDojo, however, so this result may be optimistic for the prompt.

Transfer of ranking does not make the alerts usable at a low attack rate. MON-L's native decision flags 0.900 of compromised and 0.173 of benign P0 traces, so its implied precision at 2\% prevalence is 0.096 [0.071, 0.141] on P0 and 0.073 [0.065, 0.082] on R1. MON-A's decision flags 0.883 of compromised and 0.038 of benign P0 traces, for an implied precision of 0.320 [0.203, 0.591], an interval that reaches above 0.5, but it flags 0.313 of R1 negatives and reaches only 0.054 there. Whether a native decision is usable at 2\% therefore depends on the backbone and on the benchmark. Our registered expectation, an upper bound below 0.5 on both benchmarks, held for MON-L but not for MON-A, whose P0 interval reaches above 0.5. The two backbones agree on trace flags with a Cohen's kappa of 0.69. Thresholds do not port even for an untrained detector. On P0, reading $\theta_{r80}$ from AgentDojo training folds instead of AgentDrift training positives changes MON-L's recall by $-$0.075 [$-$0.103, $-$0.049] and its FPR by $-$0.064 [$-$0.105, $-$0.033].

Judging actions separates attempted from successful hijacks better than reading content. MON-L flags 0.212 of the 4,987 unsuccessful attacks, against 0.308 for TF-IDF trained within AgentDojo (difference $-$0.096 [$-$0.186, $-$0.006]), and MON-A flags 0.063. This comparison was added after registration, because the registered comparator, TF-IDF at its transfer threshold, flags no AgentDojo trace. Finally, MON-L shares a model family with the AgentDrift generator, and its AUROC advantage over MON-A is larger on R1 than on P0 by +0.043 [+0.019, +0.067]. This suggests that familiarity helps MON-L on AgentDrift, although the backbones also differ in capability.

\section{Discussion and Recommendations}
\label{sec:discussion}

\subsection{What the Findings Mean}

\emph{Rare attacks.} For the content-blind detectors, and for the monitor's native decision with the open backbone, a high AUROC coexisted with alert streams that a low attack rate would fill with false alarms. This is the base-rate effect that Axelsson \cite{axelsson2000} described for intrusion detection, and it follows from the prevalence dependence of precision \cite{davis2006,saito2015}. The headline precision also moved threefold with the share of hard negatives, an effect that prevalence alone does not capture. Prevalence-weighted average precision makes AUPRC comparable across prevalences, but a precision figure means little unless the negative mix is stated.

\emph{Unseen tools.} The ranking reversals came mainly from inputs never seen in training. MCPShield \cite{mcpshield} showed that tool-name metadata inflates estimates under random splits. Our results add that the same features can harm a combined model once many test tools are new, that the harm grows gradually with the share of new tools, and that dropout on tool inputs largely removes it. This is a concrete and correctable form of the inversion of in-distribution and out-of-distribution rankings discussed by Teney et al. \cite{teney2023}. Thresholds are a separate problem. They also failed for SE, which has no untrained input, and for the untrained monitor, so thresholds need checking on local traffic whatever the representation.

\emph{Shortcuts.} Both benchmarks were nearly separable by surface content, through generator templates on AgentDrift and fixed attack strings on AgentDojo, and the behavioral signal that survived zero coverage rested on a length cue that the attacker writes. These are instances of shortcut learning \cite{geirhos2020}. They explain why the detectors that were best within each benchmark did not transfer, and they suggest that strong in-benchmark scores, including those of published detectors evaluated on these benchmarks, may partly reflect how the data were generated. The only detector that transferred, the released monitor, reads neither the observation text nor its length and instead judges whether each action serves the user's request. Its transfer is consistent with that judgment being closer to the behavior of interest, although our design cannot show it.

\emph{Reading a failed transfer.} The within-target control remains necessary to interpret near-chance transfer, but it supported a failure of transfer only for some representations, and the outcome for LR-C turned on small margins and on the population. Agent-Sentry \cite{agentsentry} applies a similar control to benchmarks that share agents, tools and environments, whereas our two benchmarks differ at once in label semantics, attack construction, task sources, trace origin and tool vocabulary. Our evidence therefore cannot say which difference causes the failure. For prompt classifiers, Fomin \cite{fomin2026} found that pooled cross-validation overstates leave-one-dataset-out AUC by 8.0 to 16.5 points. Our within-minus-transfer gaps of +0.263 to +0.478 AUROC are larger, but they combine several shifts that Fomin's setting keeps apart.

\subsection{Recommendations for Evaluating Trace Detectors}

\begin{enumerate}
\item Report AUROC together with prevalence-weighted average precision and precision at a fixed threshold, at stated attack prevalences such as 1\% to 10\%, with intervals from the full test pool.
\item State the negative mix behind every precision figure, and report the FPR by negative class and the recall by attack pattern.
\item Report tool coverage $\kappa$ for every split, and include a split in which many test tools are new.
\item Fix thresholds without test data and report how they behave on each test set. When test tools may be unseen, train tool-indexed inputs with dropout.
\item Include a content baseline, even a simple one, to show how far a benchmark is separable by surface content, and test cues that the attacker controls, such as observation length, by changing them.
\item Pair every cross-benchmark claim with a within-target control, report its outcome per representation and per population, and state how unsuccessful attacks are labeled.
\item Evaluate released and untrained monitors by the same protocol, including their native decision at a stated prevalence, since a monitor that ranks well can still raise mostly false alerts.
\end{enumerate}

\subsection{Benchmark Design and Deployment}

Benchmarks for trace detectors would be more informative if hard negatives shared the templates of attacked traces, if attack text were varied rather than drawn from a few fixed strings, if generation prompts and scripts were released, and if tool overlap across splits were documented. For a security team that audits agent sessions after the fact, a detector validated on one benchmark should not be assumed to work on the team's own agents and tools. Its threshold should be checked on local traffic before its alerts are trusted, and label-free recalibration can restore an alert budget but not ranking quality. An adaptive attacker could also change the cues that content features rely on, a case that we did not test and that only Praetor \cite{praetor} among the audited works evaluates.

\section{Limitations}
\label{sec:limitations}

\emph{Internal validity.} Only the monitor's plan was prospective in full (Appendix~\ref{app:plans}). Several readings were added after related results were seen, and several decisions are borderline. The GRU uses three training seeds in most settings, whose means differ from ten-seed means by at most 0.015 AUROC where both exist.

\emph{Construct validity.} We score whole trajectories after the fact and evaluate neither step-level localization nor scoring before each call. The two benchmarks encode different notions of compromise, DoS runs are labeled by task failure, and the write-like features rest on a keyword list.

\emph{External validity.} AgentDrift is synthetic, comes from one generator, is nearly separable by surface content and has fully disjoint tool sets across domains, so R3 is an extreme vocabulary shift. On recorded AgentDojo traces, a leave-one-suite-out analogue of R3 did not reproduce the C-versus-B reversal consistently. We study one benchmark pair in one direction, with a small AgentDojo benign pool (392 traces) and no adaptive attacker. SE reimplements one published architecture and is not the published detector. We include one released monitor prompt with two backbones of our choice, without its PromptGuard layer and run after the fact on recorded actions, so its results bound the released prompt and not LlamaFirewall as deployed. Its graded score is read from token probabilities that the released scanner does not use, its open backbone shares a model family with the AgentDrift generator, and Meta may have refined the prompt with AgentDojo in view. The findings may not carry over to other detectors, agent ecosystems or deployment traffic.

\section{Conclusion}

Three evaluation choices that are common in the audited literature hide failures that matter for detecting hijacked LLM agents. Balanced test sets hide that a detector with a high AUROC raises mostly false alerts when attacks are rare, and that its precision depends on the negatives as much as on prevalence. Test sets whose tools were all seen in training hide that rankings reverse as tools become new, a failure that on AgentDrift comes mainly from untrained inputs and that dropout on tool inputs largely removes, whereas thresholds fail for wider reasons. Testing on the training benchmark hides that no trained detector we evaluated ranks AgentDojo traces better than chance after training on AgentDrift, and both benchmarks are nearly separable by surface content. A released action monitor that is trained on neither benchmark does transfer, yet with its open backbone its released decision rule would still mostly raise false alerts at a 2\% attack rate. Finding which benchmark difference causes the transfer failure will require benchmark pairs that vary one factor at a time, and deciding whether monitors are usable will require evaluating them at realistic attack rates.

\section*{Code, Data and Extended Results}

The code, the derived data, the result files and every monitor call with its raw output and score are publicly available at \url{https://github.com/Mahmoud-Abbasi-svg/agent-trace-detector-eval}, together with an extended report that gives every analysis summarized here, each analysis plan with its deviations, and all supplementary tables. We use AgentDrift release 1.0 at commit \texttt{014a514f} and the AgentDojo recorded runs at commit \texttt{089ed468}. All results except those of SE and MON-L were computed on CPUs. SE ran on one node of a Slurm cluster, and MON-L on one NVIDIA H100 80GB HBM3 with vLLM 0.29.0. MON-A was queried through the provider's batch interface. Scripts generate all tables and figures from the result files.

\section*{Ethics and Data Statement}

Both benchmarks are public, AgentDrift under CC BY 4.0 and AgentDojo under the MIT license. AgentDrift is synthetic, and its documentation notes that some generated company names coincide with real organizations, although no content describes any real organization. We add no personal data. Our observation-length finding concerns simple detectors and a quantity that indirect injectors already control. We report it so that evaluations test for it, and we release no attack artifact beyond the padding procedure described here.

\section*{Declaration of Generative AI Use}

During the preparation of this work the authors used generative AI tools. After using these tools, the authors reviewed and edited the content as needed and take full responsibility for the content of the publication.

\appendices

\section{Detector Details}
\label{app:detectors}

\emph{Behavioral features.} Representation B has ten features. They are the trajectory length $m$, the number of distinct tools, tool diversity (distinct tools divided by $m$), the number of immediate repeats, the count and fraction of calls whose tool name contains one of fourteen write-like keywords (such as send, delete, transfer or share), the maximum and mean number of arguments per call, and the maximum and mean observation length in characters. An AI assistant, acting as a single non-blind annotator, marked which AgentDrift tools change state. Against that annotation the keyword flag has a precision of 0.957 and a recall of 0.440, and annotation-based features change LR-B's AUROC by at most 0.024.

\emph{Logistic regression and TF-IDF.} LR-B and LR-C use L2 regularization with inverse strength 1, the lbfgs solver, class weights inversely proportional to class frequency and features standardized with training statistics. TF-IDF uses lowercase word unigrams and bigrams with a minimum document frequency of 2, at most 50,000 features and sublinear term frequency.

\emph{GRU.} Each tool call becomes a token (0 for padding, 1 for the unknown token and one index per tool of the benign training trajectories). Tokens are embedded in 32 dimensions and passed to a one-layer GRU with 64 hidden units. Its final hidden state is concatenated with the ten standardized behavioral features and fed to a linear output layer, giving about 22k parameters. Sequences keep at most their last 16 calls. Training minimizes class-weighted binary cross-entropy with Adam \cite{kingma2015adam} (learning rate $10^{-3}$, coupled weight decay $10^{-4}$), with batch size 64, gradient clipping at norm 1.0 and at most 30 epochs. Early stopping uses the AUPRC of a stratified 10\% validation hold-out (patience 5). We use training seeds 0, 1 and 2, and ten seeds for the comparisons with SE.

\emph{SE.} SE follows the DriftNet architecture \cite{driftnet}, with frozen all-mpnet-base-v2 step embeddings \cite{mpnetmodelcard} and four world features passed to a small transformer, trained without the THOUGHT field and averaged over ten seeds. On the task-disjoint split its F1 is 0.973 against a published 0.983, in a setting that is not like for like. Zeroing its four world features changes neither its transfer nor its within-AgentDojo AUROC detectably.

\emph{Monitor.} Each step is rendered as the released demonstration renders an action, with the tool name, the arguments and an empty thought, and the AgentDojo user request is taken from the recorded run files. An output that cannot be parsed takes the released default, which flags the step. MON-L uses greedy decoding with output constrained to the released schema, and every MON-L call returned a valid output. MON-A was queried at its lowest thinking setting and with default sampling, because the batch interface rejects a temperature for this model, and its 174 refusals on P0 take the default flag. Excluding them lowers its P0 FPR from 0.038 to 0.028. A repeatability check that repeated 500 calls per backbone (492 usable for MON-A) gave a conclusion agreement of 0.97 for each.

\section{AgentDojo Extraction and Populations}
\label{app:dojo}

Removing runs with fewer than two tool calls drops 236 benign, 70 compromised and 2,260 unsuccessful-attack runs. The removal is asymmetric across classes and models. It takes, for example, 77 of 132 \texttt{gemini-2.0-flash-001} benign runs, and because trace length alone separates compromised from benign traces with an AUROC of 0.675, the filter shapes the negatives. P0 has a native prevalence of 0.854, 117 user-task identifiers and 30 injection goals. In 65 benign traces the user task was an injection task, and removing them gives population C1. The model-balanced population reweights the positives to the model mix of the benign traces, and P0F adds the 4,987 unsuccessful attacks to P0 as negatives.

An earlier version of this study paired each tool call with its observation through the tool-call identifier. Gemini runs leave that identifier empty and some GPT runs reuse it, so affected steps received another step's observation, and the earlier extractor also cut observations at 2,000 characters. The corrected extractor pairs calls and observations by position, checked by a unit test, and stores observations uncut. The fix changes 413 of the 2,686 P0 traces by observation content and 408 by per-step observation length. Every AgentDojo number here comes from the corrected extraction.

\section{Decision Rules}
\label{app:rules}

\emph{Reversal.} A reversal in an R3 fold has two requirements after Holm correction within the five folds. One detector must hold an in-distribution advantage over another that is distinguishable from zero in R1dm, and the R3 paired difference must be distinguishable from zero with the opposite sign. AUROC is the primary metric and wAUPRC at 10\% the secondary one. Threshold failure is a change in recall or FPR at $\theta_{r80}$, relative to R1dm, that is distinguishable from zero.

\emph{Cross-benchmark outcome.} Within-AgentDojo detection is reasonable when AUROC is at least 0.80 with a lower 95\% bound of at least 0.70, and wAUPRC at 10\% is at least 0.30 with a lower bound above 0.10. Transfer is near chance when its AUROC interval includes 0.5 or its AUROC is below 0.60. The second clause is one-sided and post hoc, and under it an interval lying entirely below 0.5 also counts as near chance. A failure of transfer requires reasonable within-AgentDojo detection under UT on P0 and C1, near-chance transfer and a within-minus-transfer difference distinguishable from zero. A weak representation is one whose within-AgentDojo detection under UT is not reasonable on P0 or C1 while transfer is also poor, and a population-dependent outcome is one in which within-AgentDojo detection is reasonable only on the controlled populations. Other outcomes are reported as observed. A second control that groups compromised traces by injection goal (IT) gives within-AgentDojo AUROCs of 0.645 to 0.719 for the content-blind detectors and 0.987 for TF-IDF.

\section{Analysis Plans and Pre-Registered Monitor Outcomes}
\label{app:plans}

The trained detectors follow four internal analysis plans, none of them externally registered. The cross-benchmark plan was last written after the transfer and within-AgentDojo results were known, so its one-sided near-chance clause is post hoc. The ablation plan was prospective only for the tool-count model and for the R1 and R3 results of LR-B, and the GRU plan preceded every GRU result but followed all linear results. A revision plan was written after an independent pre-submission review had reported exploratory versions of most analyses, so the revision analyses are confirmatory re-runs of analyses whose approximate outcome was known. Readings added after related results were seen include the variants without observation-length features, the label-free recalibration at 2\% prevalence, the GRU-versus-C reversal, and the GRU full-logit offsets used to compare it with SE.

The monitor's plan was registered by hash before any monitor call. Its five amendments, each hashed before any monitor result was analysed, set the budget and batch order of MON-A, corrected the number of masked calls, changed the version of the serving software, fixed the draw for the repeatability check, and added a replacement comparator for outcome M5, whose registered comparator flags no AgentDojo trace. Table~\ref{tab:prereg} lists the registered predictions and their outcomes. The extended report gives every deviation from every plan.

\begin{table}[t]
\caption{Registered Predictions for the Monitor and Their Outcomes}
\label{tab:prereg}
\centering
{\footnotesize\renewcommand{\arraystretch}{1.08}\setlength{\tabcolsep}{3pt}
\begin{tabular}{@{}l>{\raggedright\arraybackslash}p{3.65cm}>{\raggedright\arraybackslash}p{1.95cm}>{\raggedright\arraybackslash}p{1.95cm}@{}}
\toprule
 & Registered prediction & MON-L & MON-A \\
\midrule
M1 & The monitor transfers: P0 and model-balanced AUROC with lower bounds above 0.60, above every AgentDrift-trained detector & Met (0.930; $+$0.410 to $+$0.457) & Met (0.934; $+$0.414 to $+$0.460) \\
M2 & The monitor also fails to transfer: P0 AUROC interval includes 0.5 & Not met & Not met \\
M3 & The native decision is unusable at 2\%: upper bound of implied precision below 0.5 on P0 and on R1 & Met (0.096, 0.073) & Not met (0.320 [0.203, 0.591] on P0) \\
M4 & The native decision or a training threshold does not port between benchmarks & Met & Met \\
M5 & Fewer unsuccessful attacks flagged than by TF-IDF at its transfer threshold & Cannot be met (TF-IDF flags none) & Cannot be met (TF-IDF flags none) \\
M5b & As M5, against TF-IDF trained within AgentDojo (amendment 5) & Met ($-$0.096 [$-$0.186, $-$0.006]) & Met ($-$0.245 [$-$0.316, $-$0.176]) \\
M6 & Generator familiarity: MON-L $-$ MON-A larger on R1 than on P0 & \multicolumn{2}{>{\raggedright\arraybackslash}p{4.1cm}}{$+$0.043 [$+$0.019, $+$0.067], an indication only} \\
M7 & The backbones disagree on an outcome & \multicolumn{2}{>{\raggedright\arraybackslash}p{4.1cm}}{On M3 only} \\
M8 & Masking every tool name moves R1 AUROC by less than 0.05 (descriptive) & No ($-$0.091 [$-$0.109, $-$0.074]) & -- \\
\bottomrule
\end{tabular}}
\par\vspace{3pt}\begin{minipage}{\columnwidth}\footnotesize M1 to M6 and M8 are read on MON-L, and MON-A is read under M7; the MON-A column gives its values for comparison. M5b replaced the comparator of M5 by amendment 5. Values are AUROC levels and differences, implied precision at 2\% prevalence, or differences in the flag rate of unsuccessful attacks.\end{minipage}
\end{table}


\begin{thebibliography}{99}
\bibitem{greshake2023} K. Greshake, S. Abdelnabi, S. Mishra, C. Endres, T. Holz, and M. Fritz, ``Not what you've signed up for: Compromising real-world LLM-integrated applications with indirect prompt injection,'' in \emph{Proc. 16th ACM Workshop Artif. Intell. Secur. (AISec)}, Copenhagen, Denmark, 2023, pp. 79--90, doi: 10.1145/3605764.3623985.
\bibitem{zhan2024injecagent} Q. Zhan, Z. Liang, Z. Ying, and D. Kang, ``InjecAgent: Benchmarking indirect prompt injections in tool-integrated large language model agents,'' in \emph{Findings Assoc. Comput. Linguistics: ACL 2024}, Bangkok, Thailand, 2024, pp. 10471--10506, doi: 10.18653/v1/2024.findings-acl.624.
\bibitem{agentdojo} E. Debenedetti, J. Zhang, M. Balunović, L. Beurer-Kellner, M. Fischer, and F. Tramèr, ``AgentDojo: A dynamic environment to evaluate prompt injection attacks and defenses for LLM agents,'' in \emph{Proc. Adv. Neural Inf. Process. Syst. (NeurIPS), Datasets Benchmarks Track}, vol. 37, Vancouver, BC, Canada, 2024, pp. 82895--82920, doi: 10.52202/079017-2636.
\bibitem{traceaegis} J. Liu et al., ``TraceAegis: Securing LLM-based agents via hierarchical and behavioral anomaly detection,'' arXiv:2510.11203v1, Oct. 2025.
\bibitem{praetor} H. Dang, ``Enforcing benign trajectories: A behavioral firewall for structured-workflow AI agents,'' arXiv:2604.26274v1, Apr. 2026.
\bibitem{agentsentry} R. Sequeira, S. Damianakis, U. Iqbal, and K. Psounis, ``Agent-Sentry: Bounding LLM agents via execution provenance,'' arXiv:2603.22868v2, May 2026.
\bibitem{driftnet} A. Pinjari and M. P. Saint-Germain, ``DriftNet: A dual-head trajectory transformer for detecting and localizing prompt injection in LLM agents,'' arXiv:2609.10892v1, Sep. 2026.
\bibitem{trajguard} L. Advani, ``Trajectory Guard: A lightweight, sequence-aware model for real-time anomaly detection in agentic AI,'' arXiv:2601.00516v1, Jan. 2026.
\bibitem{mcpshield} S. Zavrak, ``Content-aware attack detection in LLM agent tool-call traffic: An empirical study of features, architectures, and evaluation protocols,'' arXiv:2605.11053v3, May 2026.
\bibitem{skynet} C. Zhang et al., ``Skynet: Workflow-level anomaly detection for agentic AI via semantic and structural modeling,'' arXiv:2609.06835v2, Sep. 2026.
\bibitem{trajad} Y. Liu et al., ``TrajAD: Trajectory anomaly detection for trustworthy LLM agents,'' arXiv:2602.06443v1, Feb. 2026.
\bibitem{sentinelagent} X. He, D. Wu, Y. Zhai, and K. Sun, ``SentinelAgent: Graph-based anomaly detection in multi-agent systems,'' arXiv:2505.24201v1, May 2025.
\bibitem{llamafirewall} S. Chennabasappa et al., ``LlamaFirewall: An open source guardrail system for building secure AI agents,'' arXiv:2505.03574v1, May 2025.
\bibitem{axelsson2000} S. Axelsson, ``The base-rate fallacy and the difficulty of intrusion detection,'' \emph{ACM Trans. Inf. Syst. Secur.}, vol. 3, no. 3, pp. 186--205, Aug. 2000, doi: 10.1145/357830.357849.
\bibitem{davis2006} J. Davis and M. Goadrich, ``The relationship between precision-recall and ROC curves,'' in \emph{Proc. 23rd Int. Conf. Mach. Learn. (ICML)}, Pittsburgh, PA, USA, 2006, pp. 233--240, doi: 10.1145/1143844.1143874.
\bibitem{saito2015} T. Saito and M. Rehmsmeier, ``The precision-recall plot is more informative than the ROC plot when evaluating binary classifiers on imbalanced datasets,'' \emph{PLoS ONE}, vol. 10, no. 3, Mar. 2015, Art. no. e0118432, doi: 10.1371/journal.pone.0118432.
\bibitem{sommer2010} R. Sommer and V. Paxson, ``Outside the closed world: On using machine learning for network intrusion detection,'' in \emph{Proc. IEEE Symp. Secur. Privacy (SP)}, Oakland, CA, USA, 2010, pp. 305--316, doi: 10.1109/SP.2010.25.
\bibitem{pendlebury2019} F. Pendlebury, F. Pierazzi, R. Jordaney, J. Kinder, and L. Cavallaro, ``TESSERACT: Eliminating experimental bias in malware classification across space and time,'' in \emph{Proc. 28th USENIX Secur. Symp. (USENIX Security)}, Santa Clara, CA, USA, Aug. 2019, pp. 729--746.
\bibitem{arp2022} D. Arp et al., ``Dos and don'ts of machine learning in computer security,'' in \emph{Proc. 31st USENIX Secur. Symp. (USENIX Security)}, Boston, MA, USA, Aug. 2022, pp. 3971--3988.
\bibitem{geirhos2020} R. Geirhos et al., ``Shortcut learning in deep neural networks,'' \emph{Nature Mach. Intell.}, vol. 2, no. 11, pp. 665--673, Nov. 2020, doi: 10.1038/s42256-020-00257-z.
\bibitem{fomin2026} M. Fomin, ``When benchmarks lie: Evaluating malicious prompt classifiers under true distribution shift,'' arXiv:2602.14161, 2026.
\bibitem{agentdrift} A. Pinjari and M. P. Saint-Germain, ``AgentDrift: A step-labeled benchmark of injection-hijacked LLM agent trajectories,'' arXiv:2609.06972v1, Sep. 2026.
\bibitem{zhu2025melon} K. Zhu, X. Yang, J. Wang, W. Guo, and W. Y. Wang, ``MELON: Provable defense against indirect prompt injection attacks in AI agents,'' in \emph{Proc. 42nd Int. Conf. Mach. Learn. (ICML)}, ser. PMLR, vol. 267, 2025, pp. 80310--80329.
\bibitem{chen2026tracesafe} Y.-S. Chen, S.-Y. Huang, C.-L. Yang, and Y.-N. Chen, ``TraceSafe: A systematic assessment of LLM guardrails on multi-step tool-calling trajectories,'' in \emph{Proc. 3rd Conf. Lang. Model. (COLM)}, San Francisco, CA, USA, 2026, arXiv:2604.07223.
\bibitem{li2026atbench} Y. Li et al., ``ATBench: A diverse and realistic agent trajectory benchmark for safety evaluation and diagnosis,'' in \emph{Proc. 3rd Conf. Lang. Model. (COLM)}, San Francisco, CA, USA, 2026, arXiv:2604.02022.
\bibitem{zhang2025asb} H. Zhang et al., ``Agent Security Bench (ASB): Formalizing and benchmarking attacks and defenses in LLM-based agents,'' in \emph{Proc. 13th Int. Conf. Learn. Represent. (ICLR)}, Singapore, 2025.
\bibitem{li2026agentdyn} H. Li, R. Wen, S. Shi, N. Zhang, Y. Vorobeychik, and C. Xiao, ``AgentDyn: Are your agent security defenses deployable in real-world dynamic environments?'' arXiv:2602.03117, 2026.
\bibitem{fu2025raseval} Y. Fu, X. Yuan, and D. Wang, ``RAS-Eval: A comprehensive benchmark for security evaluation of LLM agents in real-world environments,'' arXiv:2506.15253, 2025.
\bibitem{siblini2020} W. Siblini, J. Fréry, L. He-Guelton, F. Oblé, and Y.-Q. Wang, ``Master your metrics with calibration,'' in \emph{Advances in Intelligent Data Analysis XVIII (IDA 2020)}, ser. Lecture Notes in Computer Science, vol. 12080, Konstanz, Germany, 2020, pp. 457--469, doi: 10.1007/978-3-030-44584-3\_36.
\bibitem{boyd2013} K. Boyd, K. H. Eng, and C. D. Page, ``Area under the precision-recall curve: Point estimates and confidence intervals,'' in \emph{Proc. Eur. Conf. Mach. Learn. Knowl. Discovery Databases (ECML PKDD)}, ser. Lecture Notes in Computer Science, vol. 8190, 2013, pp. 451--466, doi: 10.1007/978-3-642-40994-3\_29.
\bibitem{mcdermott2024} M. B. A. McDermott, H. Zhang, L. H. Hansen, G. Angelotti, and J. Gallifant, ``A closer look at AUROC and AUPRC under class imbalance,'' in \emph{Proc. Adv. Neural Inf. Process. Syst. (NeurIPS)}, vol. 37, Vancouver, BC, Canada, 2024, pp. 44102--44163, doi: 10.52202/079017-1400.
\bibitem{barbero2022} F. Barbero, F. Pendlebury, F. Pierazzi, and L. Cavallaro, ``Transcending TRANSCEND: Revisiting malware classification in the presence of concept drift,'' in \emph{Proc. IEEE Symp. Secur. Privacy (SP)}, San Francisco, CA, USA, 2022, pp. 805--823, doi: 10.1109/SP46214.2022.9833659.
\bibitem{miller2021} J. Miller et al., ``Accuracy on the line: On the strong correlation between out-of-distribution and in-distribution generalization,'' in \emph{Proc. 38th Int. Conf. Mach. Learn. (ICML)}, ser. PMLR, vol. 139, 2021, pp. 7721--7735.
\bibitem{teney2023} D. Teney, Y. Lin, S. J. Oh, and E. Abbasnejad, ``ID and OOD performance are sometimes inversely correlated on real-world datasets,'' in \emph{Proc. Adv. Neural Inf. Process. Syst. (NeurIPS)}, vol. 36, 2023, pp. 71703--71722, doi: 10.52202/075280-3139.
\bibitem{cho2014gru} K. Cho et al., ``Learning phrase representations using RNN encoder-decoder for statistical machine translation,'' in \emph{Proc. Conf. Empirical Methods Natural Lang. Process. (EMNLP)}, Doha, Qatar, Oct. 2014, pp. 1724--1734, doi: 10.3115/v1/D14-1179.
\bibitem{salton1988tfidf} G. Salton and C. Buckley, ``Term-weighting approaches in automatic text retrieval,'' \emph{Inf. Process. Manage.}, vol. 24, no. 5, pp. 513--523, 1988, doi: 10.1016/0306-4573(88)90021-0.
\bibitem{llama3} A. Grattafiori et al., ``The Llama 3 herd of models,'' arXiv:2407.21783, Jul. 2024.
\bibitem{gptq} E. Frantar, S. Ashkboos, T. Hoefler, and D. Alistarh, ``GPTQ: Accurate post-training quantization for generative pre-trained transformers,'' in \emph{Proc. Int. Conf. Learn. Represent. (ICLR)}, 2023.
\bibitem{vllm} W. Kwon et al., ``Efficient memory management for large language model serving with PagedAttention,'' in \emph{Proc. 29th Symp. Operating Syst. Principles (SOSP)}, 2023, pp. 611--626.
\bibitem{efron1993bootstrap} B. Efron and R. J. Tibshirani, \emph{An Introduction to the Bootstrap}. New York, NY, USA: Chapman \& Hall, 1993.
\bibitem{field2007bootstrap} C. A. Field and A. H. Welsh, ``Bootstrapping clustered data,'' \emph{J. R. Stat. Soc. B, Stat. Methodol.}, vol. 69, no. 3, pp. 369--390, Jun. 2007, doi: 10.1111/j.1467-9868.2007.00593.x.
\bibitem{holm1979} S. Holm, ``A simple sequentially rejective multiple test procedure,'' \emph{Scand. J. Statist.}, vol. 6, no. 2, pp. 65--70, 1979.
\bibitem{kingma2015adam} D. P. Kingma and J. Ba, ``Adam: A method for stochastic optimization,'' in \emph{Proc. 3rd Int. Conf. Learn. Represent. (ICLR)}, San Diego, CA, USA, 2015. [Online]. Available: https://arxiv.org/abs/1412.6980
\bibitem{mpnetmodelcard} Sentence Transformers, ``all-mpnet-base-v2,'' Hugging Face model card. Accessed: Sep. 26, 2026. [Online]. Available: https://huggingface.co/sentence-transformers/all-mpnet-base-v2
\end{thebibliography}
\end{document}